\documentclass[a4paper,11pt]{article}
\usepackage[utf8]{inputenc}
\usepackage{lmodern}
\usepackage[T1]{fontenc}
\usepackage{amsmath,amsthm,amsfonts,amssymb,bm}
\usepackage{dsfont}			
\usepackage{enumitem}
\usepackage{graphicx}
\usepackage{booktabs}
\usepackage{xcolor}
\usepackage{authblk}
\usepackage[authoryear]{natbib}
\setcitestyle{square}
\usepackage[right=2.5cm,left=2.5cm,top=2cm,bottom=3cm]{geometry}
\usepackage[font={small,it}]{caption}	
\usepackage[colorlinks,linkcolor=blue,citecolor=blue,urlcolor=blue]{hyperref}
\usepackage{subfig}
\usepackage{mathtools}

\usepackage{multirow}
\usepackage{makecell}
\usepackage{afterpage}
\usepackage{footnote}

\title{ \textsc{Mid infrared spectroscopy and milk \\ quality traits: a data analysis competition \\ at the ``International Workshop on \\ Spectroscopy and Chemometrics 2021''} }
\author[1,2]{Maria Frizzarin}
\author[3]{Antonio Bevilacqua}
\author[3]{Bhaskar Dhariyal}
\author[4]{Katarina Domijan}
\author[5]{Federico Ferraccioli}
\author[6]{Elena Hayes}
\author[3]{Georgiana Ifrim}
\author[7]{Agnieszka Konkolewska}
\author[3]{Thach Le Nguyen}
\author[2]{Uche Mbaka}
\author[8]{Giovanna Ranzato}
\author[3]{Ashish Singh}
\author[9]{Marco Stefanucci}
\author[2]{Alessandro Casa\footnote{Corresponding author: School of Mathematics and Statistics, University College Dublin, Belfield, Dublin 4,
Ireland. Email: alessandro.casa@ucd.ie}}
\affil[1]{Teagasc, Animal \& Grassland Research and Innovation Centre, Moorepark, Ireland}
\affil[2]{School of Mathematics and Statistics, University College Dublin, Ireland}
\affil[3]{School of Computer Science, University College Dublin, Ireland}
\affil[4]{Department of Mathematics and Statistics, National University of Ireland, Maynooth, Ireland}
\affil[5]{Department of Statistical Sciences, University of Padova, Italy}
\affil[6]{Teagasc, Food Research Centre, Moorepark, Ireland}
\affil[7]{Teagasc, Crops Research Centre, Oak Park, Ireland}
\affil[8]{Department of Animal Medicine, Production and Health, University of Padova, Italy}
\affil[9]{Department of Economics, Business, Mathematics and Statistics, University of Trieste, Italy}
\date{}                     

\begin{document}

\maketitle

\begin{abstract}
A chemometric data analysis challenge has been arranged during the first edition of the ``International Workshop on Spectroscopy and Chemometrics'', organized by the Vistamilk SFI Research Centre and held online in April 2021. The aim of the competition was to build a calibration model in order to predict milk quality traits exploiting the information contained in mid-infrared spectra only. Three different traits have been provided, presenting heterogeneous degrees of prediction complexity thus possibly requiring trait-specific modelling choices. In this paper the different approaches adopted by the participants are outlined and the insights obtained from the analyses are critically discussed.  
\end{abstract}


\section{Introduction}
Following the interesting results obtained during similar events \citep[see e.g.][and references therein]{pierna2011case,pierna2020applicability}, a chemometric challenge has been held during the inaugural edition of the ``International Workshop on Spectroscopy and Chemometrics'', organized by the Vistamilk SFI Research Centre in April 2021. 

The dataset provided to the participants contained the values for three milk quality traits for different milk samples along with the corresponding mid-infrared spectra. Mid-infrared spectroscopy (MIRS) represents a convenient and non-disruptive way to collect vast amounts of data in a relatively cheap and fast way. In recent years, such data have been proven useful to predict several different quantities of interest in the dairy framework; see for example the results in \citet{cecchinato2009mid,mcparland2014mid,bonfatti2017comparison} for coagulation properties of milk, animal energy efficiency, and fatty acids concentration, respectively. 
Moreover, nowadays mid-infrared spectroscopy is routinely used in the analysis of cow milk samples in order to quantify fat, protein and lactose content, providing a convenient and reliable alternative to other techniques such as wet chemistry; see \citet{de2014invited} for a recent review on the various applications of MIRS in the dairy framework.
Despite these advantages, MIRS data introduce some relevant challenges from a modelling perspective, such as high-dimensionality and the strong correlations among the variables (i.e. the wavenumbers). For this reason several different strategies have been explored in literature and readers may refer to the recent work by \citet{frizzarin_predicting_2021} to see how different machine learning and statistical techniques performed on such data. 

The primary objective of the competition was, therefore, to evaluate the ability of the participants to propose different regression strategies to predict milk quality traits. In particular, it has been required to provide accurate predictions for the traits exploiting only the information contained in the corresponding milk spectra. The variables to be predicted have been chosen in order to introduce some trait specific challenges, as it will be clear in the next section. 

Different participants, or groups of, took part in the competition. In this work, the best six contributions, evaluated in terms of their predictive performances, are presented focusing both on the results obtained and on the subjective choices the participants made in terms of pre-processing and modelling paradigms.

\section{Data description and challenge}\label{sec:sec2}
A dataset containing 622 milk samples from 622 cows was collected between August 2013 and August 2014 from 7 different Irish research herds. The samples originated from Holstein-Friesian, Jersey and Norwegian Red cows, as well as their crosses; all cows were fed a predominantly grass-based diet with occasional concentrate and grass silage supplementation. The samples were collected during morning and evening milking at different stages of lactation and different parities. All samples were analyzed by the same MilkoScan FT6000 (Foss Electronic A/S, Hillerød, Denmark) milk analyzer, producing 1060 transmittance data points in the mid-infrared light region. 

\begin{figure}[t]
    \centering
    \includegraphics[scale = 0.5]{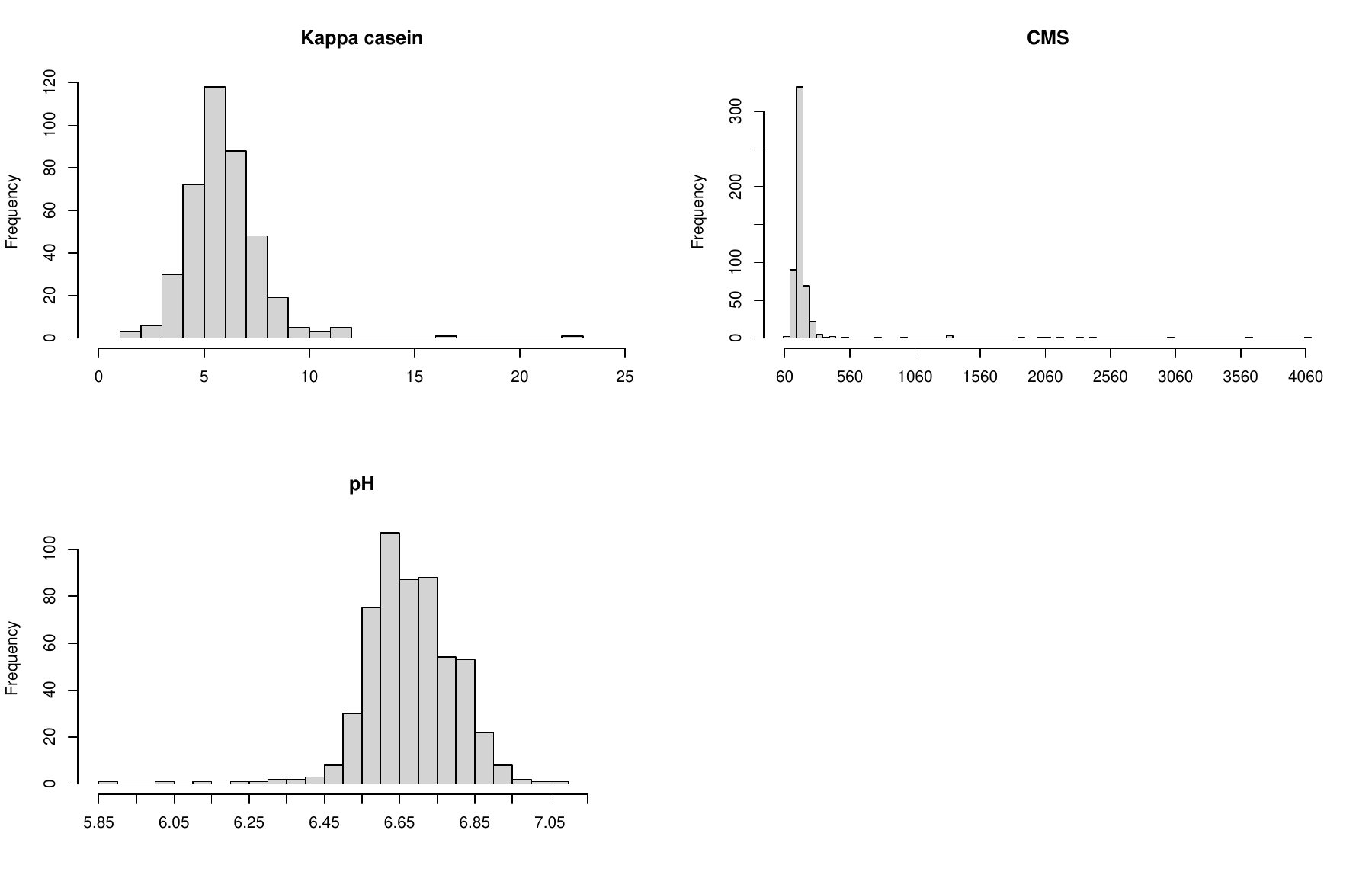}
    \caption{Frequency distributions of the three reference traits for the training set.}
    \label{fig:trait_hist}
\end{figure}

Milk pH, Casein Micelle Size (CMS) and $\kappa$-casein were provided for the analyses. Milk pH of all samples was assessed with a SevenCompact pH meter S220 (Mettler Toledo AG, Greifensee, Switzerland). The casein micelle hydrodynamic diameter was determined using a Zetasizer Nano system (Malvern Instruments Inc., Worcester, UK). Lastly, $\kappa$-casein was determined using reverse-phase high performance liquid chromatography (HPLC) using an adaptation of the method of Visser et al. (1991) and is expressed as grams per liter of milk. 

The original dataset was divided into training and test set. The training set contained 399 observations for $\kappa$-casein, 538 observation for CMS, and 548 observations for pH. Some spectra were therefore associated with only one or two of the reference traits, with the missing trait being considered as a missing value. Some summary statistics for the mentioned traits are reported in Table \ref{table:descriptive_stat} while a graphical representation of their frequency distributions is provided in Figure \ref{fig:trait_hist}. For all the observations the corresponding mid-infrared spectra, with measures for 1060 wavenumbers, were available. A test dataset has been provided to the participants, with 69 spectra with unknown values for the three reference traits. Therefore, the objective of the competition was to predict the test set values for $\kappa$-casein, CMS, and pH values by exploiting the spectral information only. Note that the information about the specific wavenumbers has not been shared with the participants; nonetheless they were aware that the region spanned by the provided spectra goes from 900 to 5000 cm$^{-1}$.
The three traits provided have been carefully selected according to their different characteristics \citep[see][]{frizzarin_predicting_2021}; CMS has been shown to have low predictability while $\kappa$-casein and pH has a medium-low and a medium-high predictability, respectively. Moreover, all these traits are important for milk processing. The datasets provided for the competition are publicly available at \texttt{http://hdl.handle.net/11019/2597}.

The contributions from the participants have been ranked considering the \emph{Root Mean Squared Error} computed on the test dataset (RMSEP)
$$
\text{RMSEP}_{kj} = \sqrt{\frac{\sum_{i=1}^n (y_{ik} - \hat{y}^{(j)}_{ik})^2}{n}}
$$
where $y_{ik}$ and $\hat{y}^{(j)}_{ik}$ represents respectively the $i$-th reference value and the corresponding prediction for the $k$-th trait by the $j$-th participant, while $n$ is the total number of observations in the test set. In order to provide a single measure to rank the performances across all the traits a \emph{relative error} has been considered, that has been defined as 
\begin{eqnarray}\label{eq:rel_err}
\text{RERR}_j &=& \frac{1}{3}\left(\sum_{k=1}^3 \frac{\text{RMSEP}_{kj}}{\text{min}_j \,\, \text{RMSEP}_{kj}} \right)
\end{eqnarray}
with $j$ spanning over the total number of participants. Thus for each trait, each participant’s RMSEP is divided by 
the best RMSEP for that trait and this is summed up over the three traits.  The more the value of (\ref{eq:rel_err}) is close to 1, the better the corresponding method has performed over the available traits. The rationale behind the choice to consider (\ref{eq:rel_err}) to evaluate the results, lies in the necessity to have a relative measure being trait independent and not being influenced by the different error magnitudes and degree of difficulty for the three traits.

\begin{table}[t]
\centering
\caption{Mean, median, standard deviation (SD), minimum and maximum values for the provided traits.}
\begin{tabular}{l|ccccc}
  \hline

  Trait & Mean & Median & SD & Minimum & Maximum \\ 
  \hline
  $\kappa$-casein & 5.92 & 5.71 & 1.83 & 1.36 & 22.23 \\
  CMS & 229.99 & 172.45 & 343.10 & 63.12 & 4063.00 \\
  pH & 6.68 & 6.68 & 0.12 & 5.86 & 7.06 \\
   \hline
\end{tabular}
\label{table:descriptive_stat}
\end{table}

\section{Modelling approaches and results}\label{sec:sec3}

\subsection{Participant 1}\label{sec:sec3_1}
The data have been analyzed by considering different \emph{tabular methods} which, in the machine learning literature, are considered in those situations where it is possible to represent the data in terms of arrays or tables. As a consequence, the ordering of the wavelengths, and their possible chemical relations due to their spectral proximity, has not been taken explicitly into account. All the analyses reported have been conducted using pandas, sklearn and matplotlib libraries \citep[see][and references therein for the considered methods]{pedregosa2011scikit} in Python \citep{python} and the code required to reproduce the results is freely available at \texttt{https://github.com/mlgig/vistamilk-spectroscopy-challenge}.

The data analysis was performed according to a multi-step procedure. In the first exploratory step, some descriptive statistics have been computed to check the presence of missing values. Rows with missing values in the training set for the given reference traits have been removed. Possible outliers have been identified and removed coherently with the work by \citet{frizzarin_predicting_2021} which suggested to keep only those rows with a value that falls within three standard deviations from the mean of the reference trait. Moreover, data visualization tools have been exploited to see whether there were unusual values with respect to the normal expected range, and to understand the behaviour of the samples for each one of the reference traits. 

\begin{savenotes}
\begin{table}[b]
\caption{Different tabular methods considered, listed according to their sklearn implementations.}
    \centering
    \begin{tabular}{|l|l|}
         \hline
         \textbf{Modeling Strategy} &  \textbf{Implementation in sklearn}\\ \hline
         Linear Models & LinearRegression(), PLSRegression(), RidgeCV() \\ 
                        & Lasso(), ElasticNet(), SVR(kernel=`linear')\\\hline
         Ensembles Methods & RandomForestRegressor(n\_estimators=100), \\
                        & GradientBoostingRegressor(n\_estimators=100) \\
                        & Other variants: Xgboost, LightGBM \\ \hline
         Other Approaches & KNeighborsRegressor(n\_neighbors=1), SVR(kernel='rbf'), \\    
         & MLPRegressor(), FCN, Resnet \citep{chollet2015keras}. \\\hline
    \end{tabular}
    \label{tab:tabular-models}
\end{table}
\end{savenotes}

In the subsequent step, methods belonging to different categories and learning strategies have been selected. Simpler models, being easier to interpret in terms of the domain knowledge, were preferred. Regularized models were mainly considered in order to deal with noisy and correlated features, and more complex feature selection or dimensionality reduction methods have not been tested. Some of the models evaluated are listed in Table \ref{tab:tabular-models}. When running the analyses a 4-fold cross-validation (CV) scheme was adopted and the cross-validation RMSE (RMSECV) was employed to compare the performances of the different algorithms. Additionally, a single train-test split was also run to check what the models learned and, when possible, which were the important features. The CV results were compared to the single-split results. Once the best model for each trait was identified based on the performances from CV and single-split, that model was used to predict the unlabeled test set.

\begin{table}[t]
    \caption{RMSECV averaged over the four cross-validation folds (standard deviations in brackets).}
    \centering
    \begin{tabular}{l|ccc}
\hline    
Method & $\kappa$ -casein  & CMS & pH \\ \hline
LASSO & 1.51 (0.04) & \textbf{56.78 (19.35)}  & 0.12 (0.02) \\
PLSR  & 1.24 (0.02) & 58.99 (20.68) & 0.10 (0.01) \\ 
Random Forest & 1.18 (0.08) & 63.33 (17.74) & 0.09 (0.01) \\
Gradient Boosting Regressor   & 1.19 (0.1) & 64.99 (18.76) & 0.09 (0.01)  \\
SVR & 1.36 (0.05) & 57.46 (19.61) & 0.10 (0.01) \\
MLP Regressor & 1.38 (0.09) & 58.25 (18.32) & 0.18 (0.02) \\
Ridge regression & \textbf{1.16 (0.06)} & 57.16 (20.21) & \textbf{0.08 (0.01)} \\
\hline
    \end{tabular}
    \label{tab:results_tabularGeorgiana}
\end{table}

\begin{figure}[t]
    \centering
    \includegraphics[width=17cm]{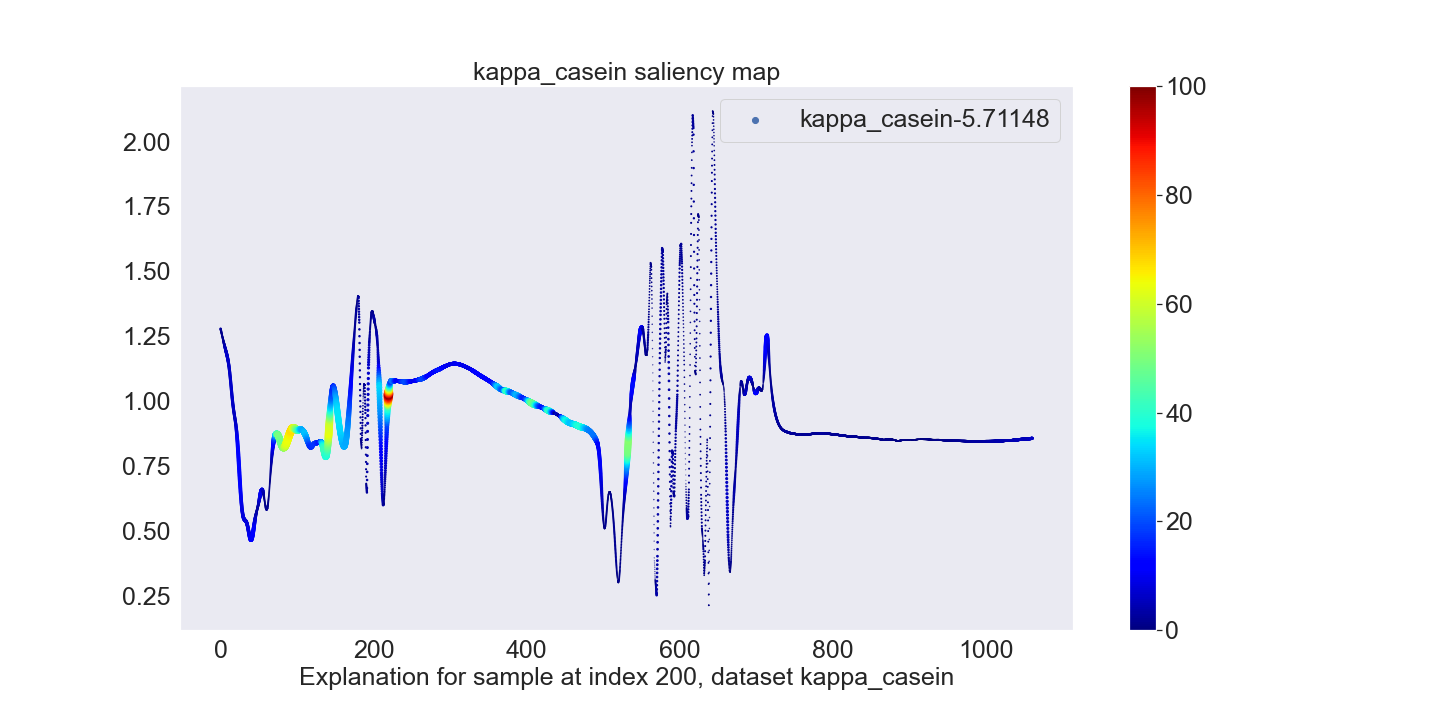}
    \caption{Saliency map computed from ridge regression estimated coefficients on a spectrum used for $\kappa$-casein prediction. High red intensities corresponds to more relevant wavenumbers.}
    \label{fig:sample-kappacasein-ridgecv-saliencymap}
\end{figure}

In Table \ref{tab:results_tabularGeorgiana} the results in terms of the RMSECV, averaged over the 4-folds, are reported for the more promising models. Interestingly, the table shows how linear models, such as PLSR and Ridge regression, provide comparable results and they have better accuracy with respect to more complex non-linear strategies. This behavior confirms what has been seen also in \citet{frizzarin_predicting_2021}, thus possibly indicating that the relations between the three traits and the spectra is close to linear. \\
For two out of three reference traits ridge regression had the best results, with the LASSO slightly outperforming it for CMS. This method provides small improvements over the three traits with respect to Partial Least Square Regression (PLSR) which is usually considered as the baseline model in the milk spectroscopy literature. Ridge regression has been run on a normalized version of the data with the penalty being tuned via an additional cross-validation scheme, with \texttt{alphas=np.logspace(-2, 2, 10)}. Moreover it is worth noting that ridge regression allows, by looking at the estimated coefficients, to identify the wavenumbers deemed as most relevant for the prediction of a specific trait. In Figure \ref{fig:sample-kappacasein-ridgecv-saliencymap} it is represented a spectrum with the ridge learned coefficients, for $\kappa$-casein prediction, highlighted as a saliency map. 

Lastly note that other strategies were tested. In fact, the wavenumbers were transformed from transmittance to absorbance and, moreover, separate analyses have been conducted with all the traits being log-transformed. Nonetheless, none of these transformations have led to improvements in the RMSECV. Given the above considerations, the ridge regression model has been used to predict the values of the traits for the test dataset.

\subsection{Participant 2}\label{sec:sec3_2}
Data collected via mid-infrared spectroscopy techniques, even if somehow resembling time series data, do not have a time component. Nonetheless, spectroscopy data have been successfully modelled as time series in the machine learning community \citep[see e.g.][]{bagnall16bakeoff,nguyen19interpretable}. Such approaches have been tested with considerable success, especially for classification, but most of these methods can be easily adapted to the regression framework \citep{Tan2020TSER}. Therefore, for the data analysis step, the following time series models were evaluated: ROCKET \citep{dempster2019rocket}, MiniROCKET, EnsembleMiniROCKET, Fully Convolutional Neural Networks (FCN), Residual Networks \citep[ResNet,][]{IsmailFawaz2018deep} and the recently proposed SeQuence Miner on Multiple Representation of time series \citep[MrSQM,][]{nguyen2021mrsqm}. 


ROCKET was originally proposed as a time series classifier that uses multiple random convolutional kernels to extract patterns in the time series. A random kernel is a vector of random weights and is normally shorter than the input time series. The convolution computes the dot product between the kernel itself and a series of windows (of the kernel's length) sliding from the beginning to the end of the input time series. Typically, the output of the dot product is greater at the window where the kernel matches the input. The output then can be used to produce features (e.g. max pooling) for model training with ridge regression to classify time series. As a consequence the method can be easily adapted to regression problems. MiniROCKET represents an updated version of this approach which, by optimizing the choices of kernels, results in an extremely fast algorithm. In the area of time series classification, this method is currently considered to be the state of the art. A recent study demonstrated that ROCKET-type algorithms work very well on time series regression \citep[see][]{Tan2020TSER}.

A new ensemble approach, named EnsembleMiniROCKET and based on MiniROCKET model, has been specifically proposed for this data challenge. It consists of an ensemble of 100 MiniROCKET regression models run on subsets of adjacent wavenumbers. The subsets are chosen randomly and they have varying dimensions, spanning from 100 to 250 features.

Other methods proposed in the deep learning literature, such as the FCN and the ResNet architectures, were tested. In addition MrSQM, a time series classification method based on symbolic representations, was also evaluated. This method allows to interpret the obtained results by providing an indication about which variables are regarded as more important for prediction purposes.

Prior to the analysis, the same pre-processing steps as in Section \ref{sec:sec3_1} have been performed. The 4-fold RMSECV was used to compare the models and the results are reported in Table \ref{tab:ts_results}. The ROCKET algorithm and its modifications tend to perform better for all the three traits. Notably the ensemble approach, specifically proposed for this challenge, provided slight improvements for two out of three traits, with respect to all the considered competitors, thus giving a first indication about the soundness of the strategy and the need to test it thoroughly on different datasets. Given these considerations, EnsembleMiniROCKET has been used to provide the predictions for the unknown traits values on the test set. Lastly, after the competition some of the considered methods have been tested without considering the water absorption regions, with slight improvements in the results. 


\begin{table}[]
    \centering
    \caption{RMSECV averaged over the four cross-validation folds (standard deviations in brackets).}
    \begin{tabular}{l|ccc}
\hline    
Method              & $\kappa$-casein  & CMS & pH \\ \hline
ROCKET             & 1.24 (0.08) & {\bf 56.88 (19.06)}  & 0.10 (0.01) \\
MiniROCKET          & 1.21 (0.12) & 57.18 (19.03) & 0.09 (0.01) \\ 
EnsembleMiniROCKET  & {\bf 1.19 (0.11)} & 56.99 (18.86) & {\bf 0.08 (0.01)} \\
FCN                 & 1.35 (0.15) & 65.55 (12.40) & 0.14 (0.01) \\
ResNet              & 1.33 (0.14) & 63.71 (14.96) & 0.13 (0.01) \\
MrSQM               & 1.40 (0.11) & 56.64 (18.46) & 0.10 (0.01) \\
\hline
    \end{tabular}
    \label{tab:ts_results}
\end{table}

\subsection{Participant 3}\label{sec:sec3_3}
The data cleaning and pre-processing steps started with the identification of possible outliers. As it has been done in \cite{frizzarin_predicting_2021}, the outliers have been considered to be those values of the three traits that fell outside of 3 standard deviations from the respective mean. These outliers, along with the missing values, were removed from the training set: as a result $\kappa$-casein had 396 observations while CMS and pH had respectively 526 and 542 observations. As visible from Figure \ref{fig:trait_hist}, the distribution of CMS resulted to be positively skewed, therefore its values were transformed on a logarithmic scale for the subsequent analyses.

One of the more troublesome aspects of spectroscopy data is usually represented by their high-dimensionality which is often paired with strong multi-collinearities among the wavenumbers. In order to deal with these  challenges, Principal Component Analysis \citep[PCA, see e.g.][]{wold1987principal} has been considered. More specifically, the first 30 principal components, explaining more than 90\% of the original variability, were retained. As a consequence, the training set considered in the analyses consists in the values for the traits to be predicted along with $p=30$ predictors. In order to provide coherent predictions, the testing set was transformed accordingly.  

\begin{figure}[t]
\centering
\includegraphics[scale = 0.38]{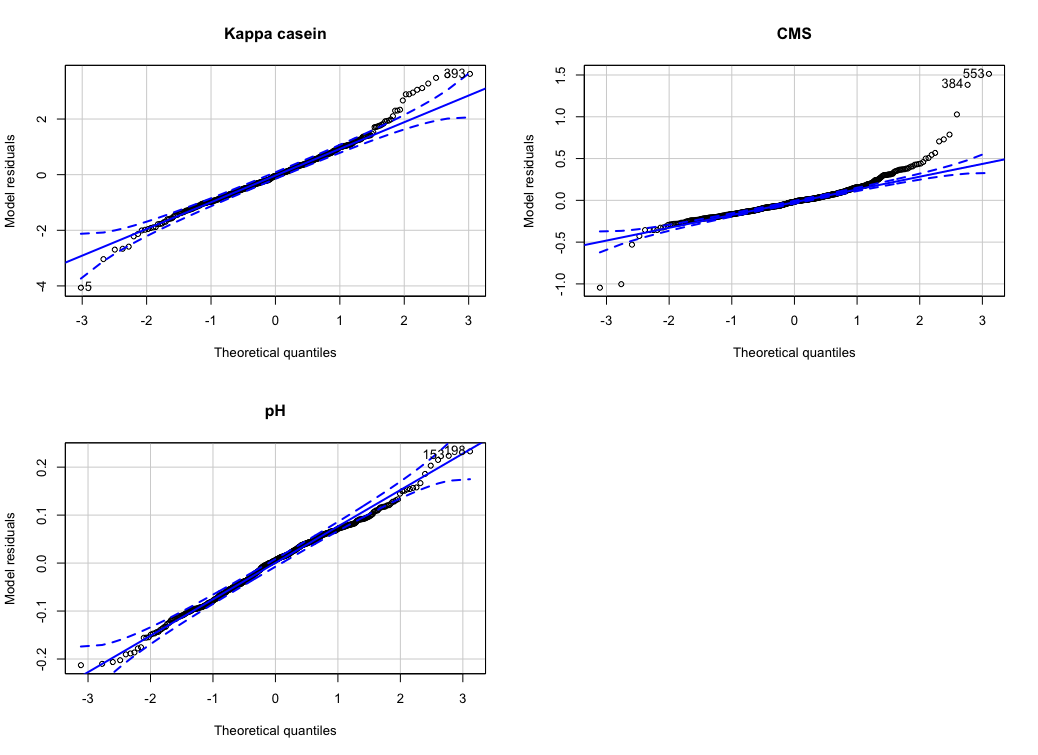}
\caption{Normal QQ-plots of the residuals of the model for $\kappa$-casein, CMS and pH. } \label{fig:fig1Gio}
\end{figure} 

As a regression strategy to predict the values for the reference traits, Generalized Additive Models have been considered \citep[see][for an overview]{friedman2001elements}. These models can be expressed as follows 
$$ \mathbb{E}(y|x_1,...,x_p)=\beta_0+\sum_{j=1}^{p}f_j(x_j)$$
where $x_1,\dots,x_p$ denote the predictors, $y$ is the outcome to be predicted and $f_j$'s are nonparametric and unspecified smooth functions. In the reported analyses, smoothing splines over the principal components have been considered for $f_j$; this might lead to a gain in terms of flexibility when describing the relationship between the trait and each single predictor. Note that similar spline-based approaches, considering PLS instead of PCA, have been already fruitfully adopted in the chemometrics framework and it might have been interesting to test them on the dataset provided \citep[see e.g.][]{durand2001local,kramer2008penalized}.\\ 
The models were built in a step-wise fashion, considering different alternatives for each single $f_j$ \citep[see][for more details]{hastie2017generalized}. More specifically, for example, $x_1$ could enter in the model either linearly or modelled by means of a spline with 2, 3 or 4 equivalent degrees of freedom. Different combinations corresponding to the possible choices have been tested and the best one was selected according to the Akaike information criterion (AIC).

Figure \ref{fig:fig1Gio} shows the normal QQ-plots for the residuals of the three models, one for each reference trait. They tend to follow a normal distribution, especially for the pH, with some discrepancies on the tails for the other two traits. The plots served as a partial confirmation about the soundness of the adopted modelling strategy which consequently has been considered to predict the trait values for the testing set. All the analyses were carried out using the \texttt{R} software \citep{RCoreTeam2020}.

\subsection{Participant 4}\label{sec:sec3_4}
Some preliminary pre-processing 
step were conducted before proceeding to the analysis. At first, 
the samples for which the response variable were not available were removed. The variables CMS and $\kappa$-casein were logarithm transformed, in order to reduce the skewness. Finally, the response variable and the spectra were both centered.

\begin{figure}[t]
    \centering
    \begin{minipage}{.5\textwidth}
        \textsf{\hspace{1cm}Regular \hspace{2.5cm} Adaptive}\par\medskip
        \centering
        \includegraphics[width=1\linewidth]{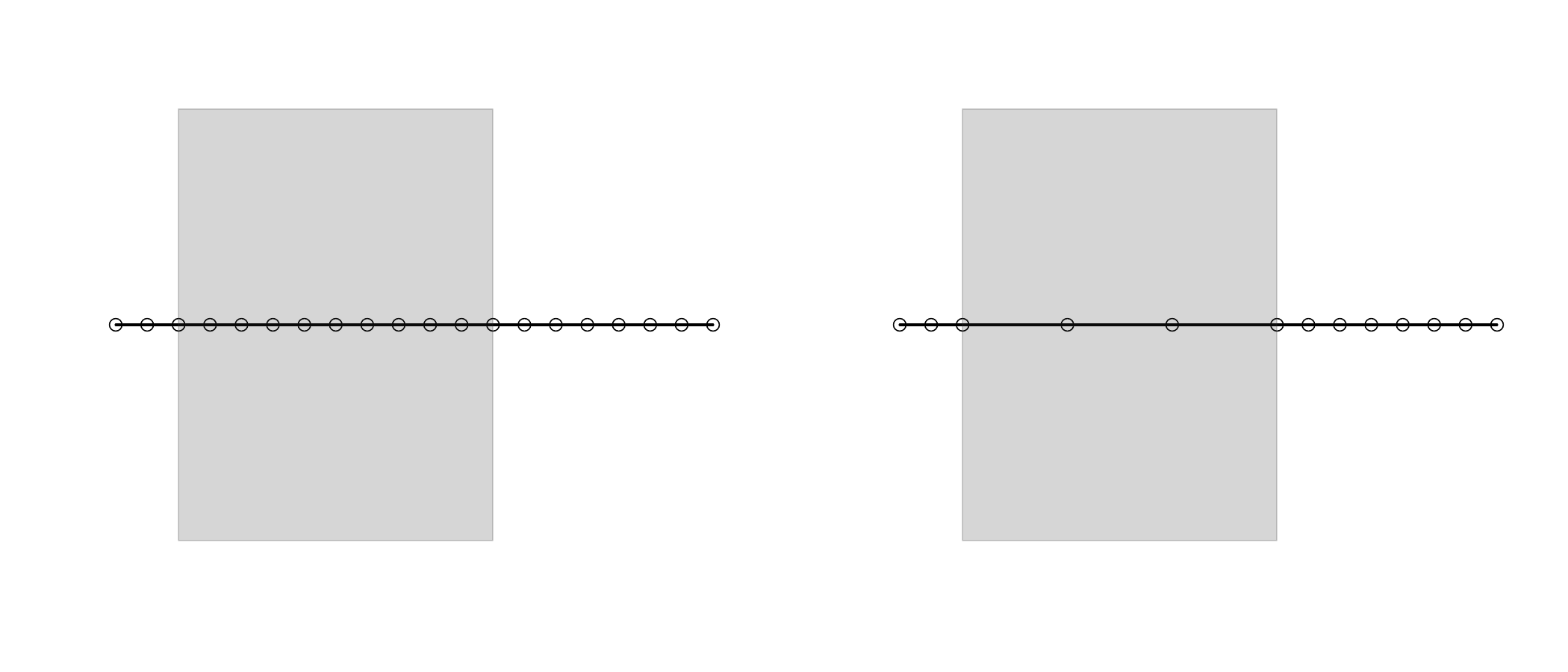} 
    \end{minipage}%
    \begin{minipage}{.5\textwidth}
        \centering
        \includegraphics[width=\linewidth]{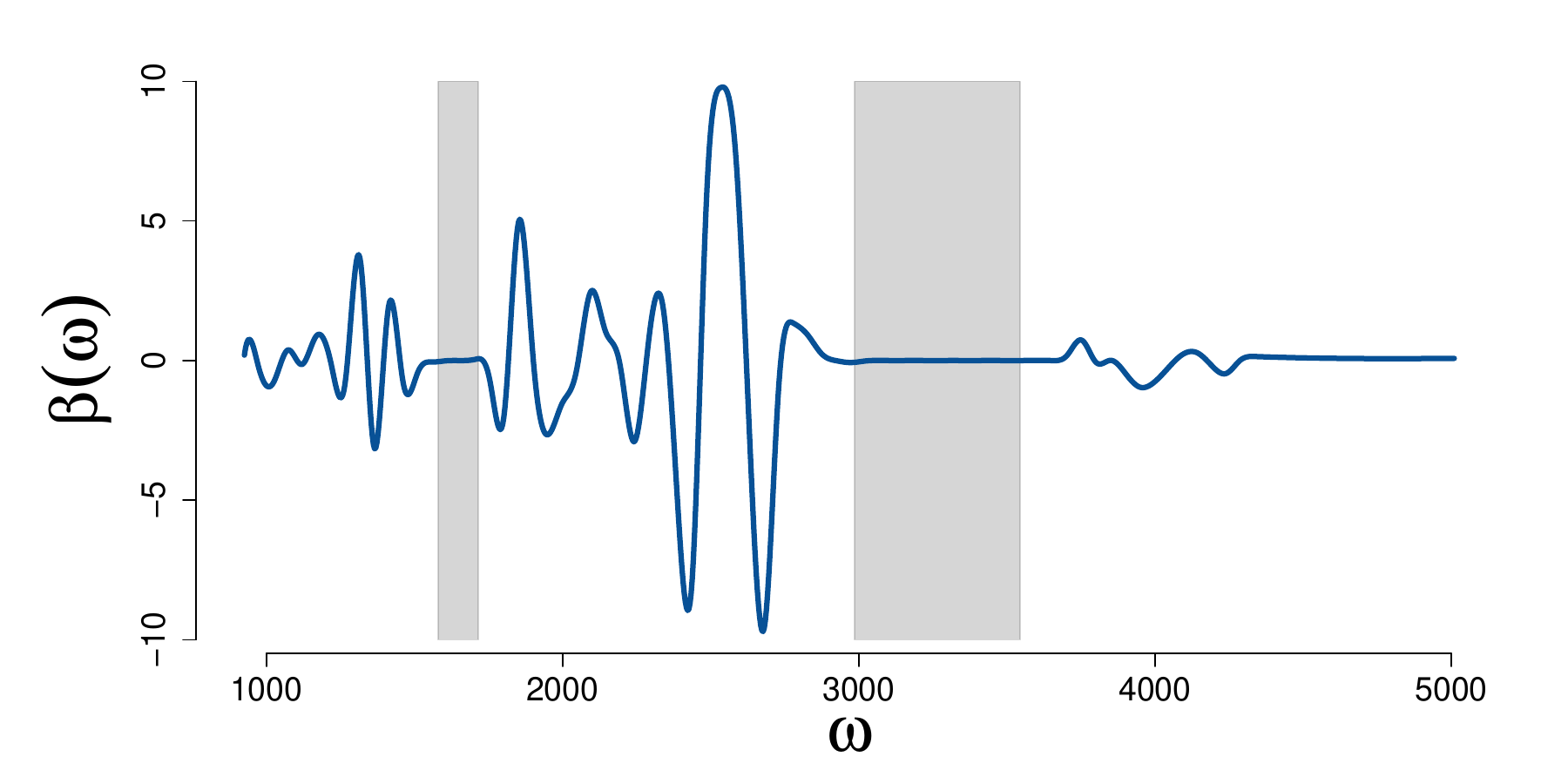}
    \end{minipage}
    \caption{Left: comparison between standard and adaptive. Right: estimated functional coefficients for the variable pH.}
    \label{fig:adapt}
\end{figure}

Since the covariates were in the form of curves, the problem was tackled with a functional regression approach \citep{FDA}, that is 
\begin{equation*}
    y = \int \beta(\omega) X(\omega) d\omega + \varepsilon,
\end{equation*}
with the response variable $y$ being in turn pH, $\kappa$--casein and casein micelle size, $X(\omega)$ the MIR spectrum and $\beta(\omega)$ the unknown functional coefficient. A B-spline basis \citep{de1978practical} defined on an adaptive grid was used, 
thus leading to different number of basis functions in different regions of the spectrum. The left panel of Figure \ref{fig:adapt} highlights the difference between a regular grid, on the left, and the adaptive grid, on the right. In particular, the adaptive grid allowed to leverage on prior information about the chemical characteristics of the spectra, for instance by reducing the number of nodes in the noisy areas induced by water absorption. These regions, highlighted in grey in the right panel of Figure \ref{fig:adapt}, corresponds to the wavenumbers from $1580$ to $1715 \, \text{cm}^{-1}$ and from $2986$ to $3545\, \text{cm}^{-1}$ \citep{grelet2015standardization}. Both ridge and lasso penalization have been used on the coefficients of the basis, in order to further shrink them in the noisy regions, deemed as less informative. Finally, a 10-fold cross-validation was used to select the best type of regularization, the number of basis and the best smoothing parameter for each of the three response variables.

The right panel of Figure \ref{fig:adapt} shows the estimated functional coefficients for the variable pH. In this particular case the selected model corresponded to the one with 75 adaptive basis and considering the lasso as regularization strategy. In the grey areas, that coincides with the noisy areas induced by water absorption, the model shrinks the coefficients to zero. This shows the strengths of the penalization combined with an adaptive functional approach. On one side, the functional nature of the data was exploited without the need to remove parts of the spectrum; on the other side, the model remained very flexible and was still able to shrink the coefficients of the non-informative regions. Moreover, the functional coefficients can be interpreted in terms of the chemical characteristics of the milk. Different parts of the spectrum are indeed linked to different chemical compositions and chemical bonds, due to the differences in light absorption \citep{grelet2015standardization}. This peculiarity can be used by experts to identify statistically significant areas of the spectrum and to make inference on their influence on the response variable. 

Due to space limitations, the results for the other two traits are not shown. In the case of $\kappa$-casein, the best model was again the lasso with 50 adaptive basis. The behavior of the coefficients was similar to the one obtained for pH in the two water regions, while the peaks differed in locations and size in the other parts of the spectrum. In the case of CMS, the best model had 15 basis and employed the ridge as regularization strategy. The results were slightly different, in particular the water regions were not zero and the overall functional coefficient presented a very smooth behavior. Nonetheless, these results might be due to the difficulty in estimating such variable and need to be discussed with an expert.

\subsection{Participant 5}\label{sec:sec3_5}


In the pre-processing steps, the data in the training set were centered and scaled, with the same transformation being applied to the test set. Moreover, CMS was log transformed, in order to reduce its skewness. Note that no outlier observations were removed and that the full set of 1060 wavenumbers was used for training and prediction. 

\begin{figure}[t]
    \centering
    \includegraphics[width=11cm, height = 7.5cm]{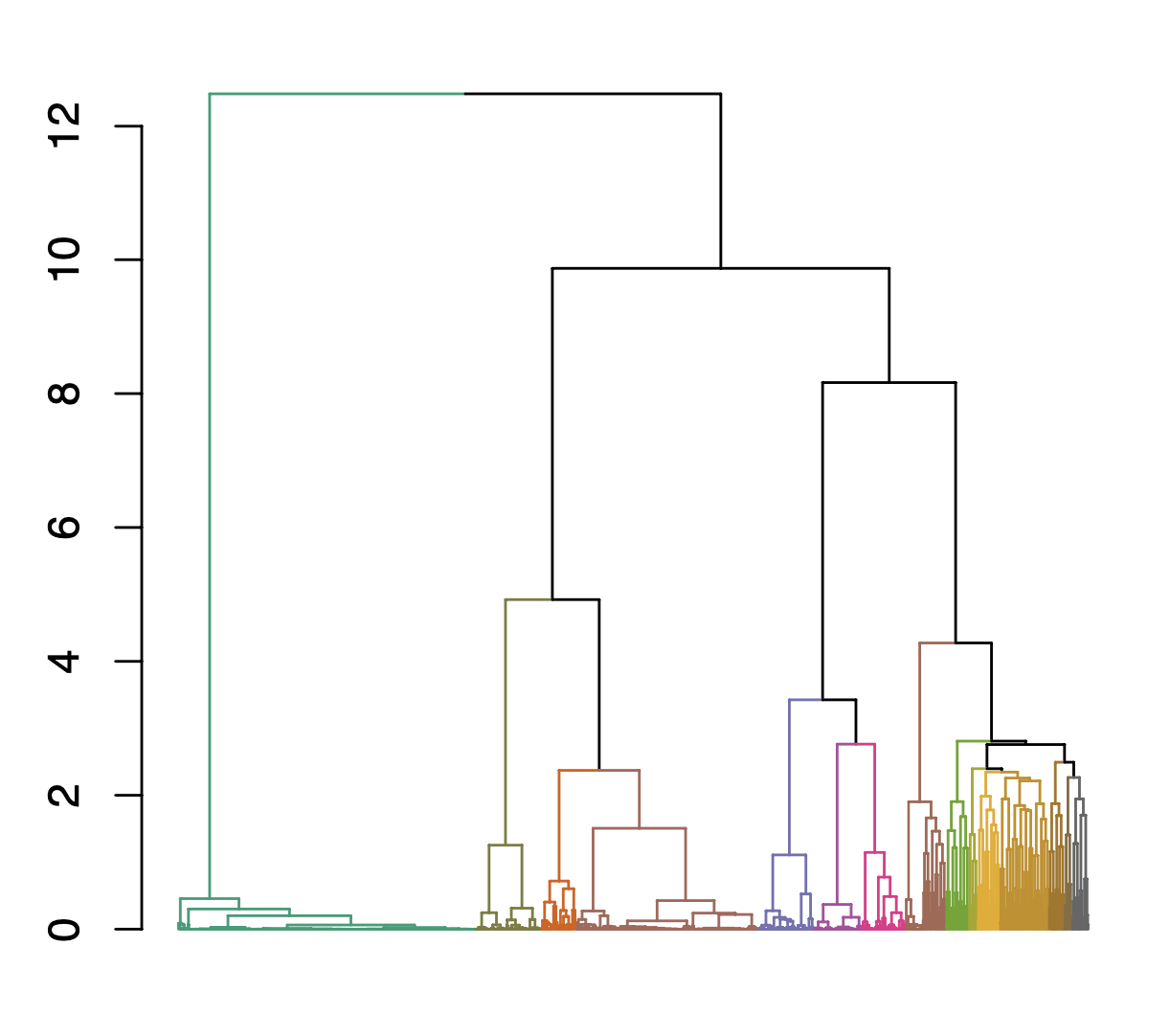}
    \caption{Dendrogram of the spectra clustered by hierarchical clustering with distance matrix based on the absolute value of correlation between the spectra and Ward's clustering criterion.}
    \label{fig:spectradendro}
\end{figure}

Hierarchical clustering with distance matrix based on the absolute value of correlation between the spectra and Ward's clustering criterion was used to find groupings of the spectra. The resulting dendrogram is displayed in Figure \ref{fig:spectradendro}; after a graphical inspection, the wavelengths were assigned to one of the fifteen clusters. The clustering was used for one of the modelling approaches described below.

The available training dataset with known trait values was initially split into 50 random splits of training (50\%) and validation (50\%) sets to select the optimal algorithm for regression. The algorithms were tuned using further cross-validation of the training subsets. 
All analysis was done in \texttt{R} and the code is available at \texttt{https://github.com/domijan/vistamilk}.
The following algorithms were used for regression:
\begin{itemize}
\item Lasso from library \texttt{glmnet} \citep{glmnet}.
\item Random forest (RF) from library \texttt{ranger} \citep{ranger} with and without regularization (RF + VS) \citep{regrf}.
\item Linear regression with the first six principal components (lm+PCA) as input features.
\item Linear regression with the first six kernel principal components (lm+kPCA) using Gaussian kernel implemented in library \texttt{kernlab} \citep{kernlab}.
\item Random forest with six kernel principal components (RF+kPCA) as input features.
\item Random forest with a Gaussian kernel in place of the transmittance values (RF+kernel) as input features.
\item Support vector machine (SVM) for regression implemented in library \texttt{e1071} \citep{e1071}.
\item Partial least squares (PLS) with 3, 4 and 5 components implemented in library \texttt{pls} \citep{pls}.
\item Linear regression with a naive approach to feature selection (lm+15); fifteen wavelengths were selected as predictors using the hierarchical clustering. From each one of the fifteen clusters in Figure \ref{fig:spectradendro}, the wavelength with the highest correlation with the response variable was selected and employed as an input to the regression model.
\item Bayesian additive regression trees \citep[BART,][]{bart} as in \texttt{bartMachine} library \citep{bartMachine}.
\item Post-hoc ensemble model that averaged over test set predictions from all of the models described above.
\end{itemize}

\begin{table}[t]
\centering
\caption{RMSE obtained over 50 random splits of the training set into training and validation for $\kappa$-casein, CMS  and pH. In brackets the standard deviations are reported.}
\begin{tabular}{l|ccc}
  \hline
  Algorithm & $\kappa$-casein & CMS & pH \\ 
  \hline
ensemble & 1.48 (0.21) & 367.5 (31.7) & 0.11 (<0.01) \\ 
lasso & 1.53 (0.21) & 368.3 (31.3) & 0.12  (<0.01) \\ 
lm+15 & 1.52 (0.18) & 366.2 (31.9) & 0.11  (0.01) \\ 
lm+PCA & 1.50 (0.20) & 367.9 (31.4) & 0.12  (<0.01) \\ 
lm+kPCA & 1.57 (0.19) & 368.2 (31.4) & 0.11  (<0.01) \\ 
PLS  & 1.50 (0.19) & 366.0 (31.4) & 0.12  (<0.01) \\ 
BART & 1.62 (0.12) & 368.3 (31.9) & 0.12 (0.01) \\ 
RF & 1.53 (0.18) & 367.9 (31.5) & 0.12  (0.01) \\ 
RF+VS & 1.52 (0.19) & 367.8 (31.6) & 0.12  (0.01) \\ 
RF+kPCA & 1.57 (0.21) & 369.4 (31.6) & 0.12  (<0.01) \\ 
RF+kernel & 1.53 (0.21) & 370.0 (31.6) & 0.12  (<0.01) \\ 
SVM & 1.54 (0.21) & 371.0 (31.3) & 0.11 (<0.01) \\ 
   \hline
\end{tabular}
\label{table:KD}
\end{table}

The predictive ability of the algorithms was evaluated according to the RMSE computed on the validation set. Table \ref{table:KD} displays the average RMSE over the 50 random splits with standard deviation in brackets. All the models considered had comparable performance; the best scoring algorithm for all three traits was the ensemble model, which slightly outperformed the other models. As a consequence, it was selected and trained over the entire training set and used to predict the values for the test set. 


\subsection{Participant 6}\label{sec:sec3_7}
First of all, an exploratory data analysis has been performed with Unscrambler X (Camo Analytics) and with the \texttt{R} software. The data were visualized by plotting the raw spectra, as well as the calculated interquartile-range of each wavelength which allowed to identify two highly noisy regions. Several spectral pre-treatments were tested such as Standard Normal Variate, first and second derivative Savitzky-Golay transformations and Multiplicative Scatter Correction (MSC) with the latter one being chosen as the most promising. According to the identification of the high-noise-level regions, in the subsequent analyses two approaches have been taken: in the first case the original 1060 wavenumbers have been considered while, in the latter, the noisy regions were removed leading to datasets with 533 features. In both the cases, CMS values were log-transformed to obtain data following more closely a Gaussian distribution.

\begin{figure}[t]
    \centering
    \includegraphics[width = 16.3cm, height = 7.5cm]{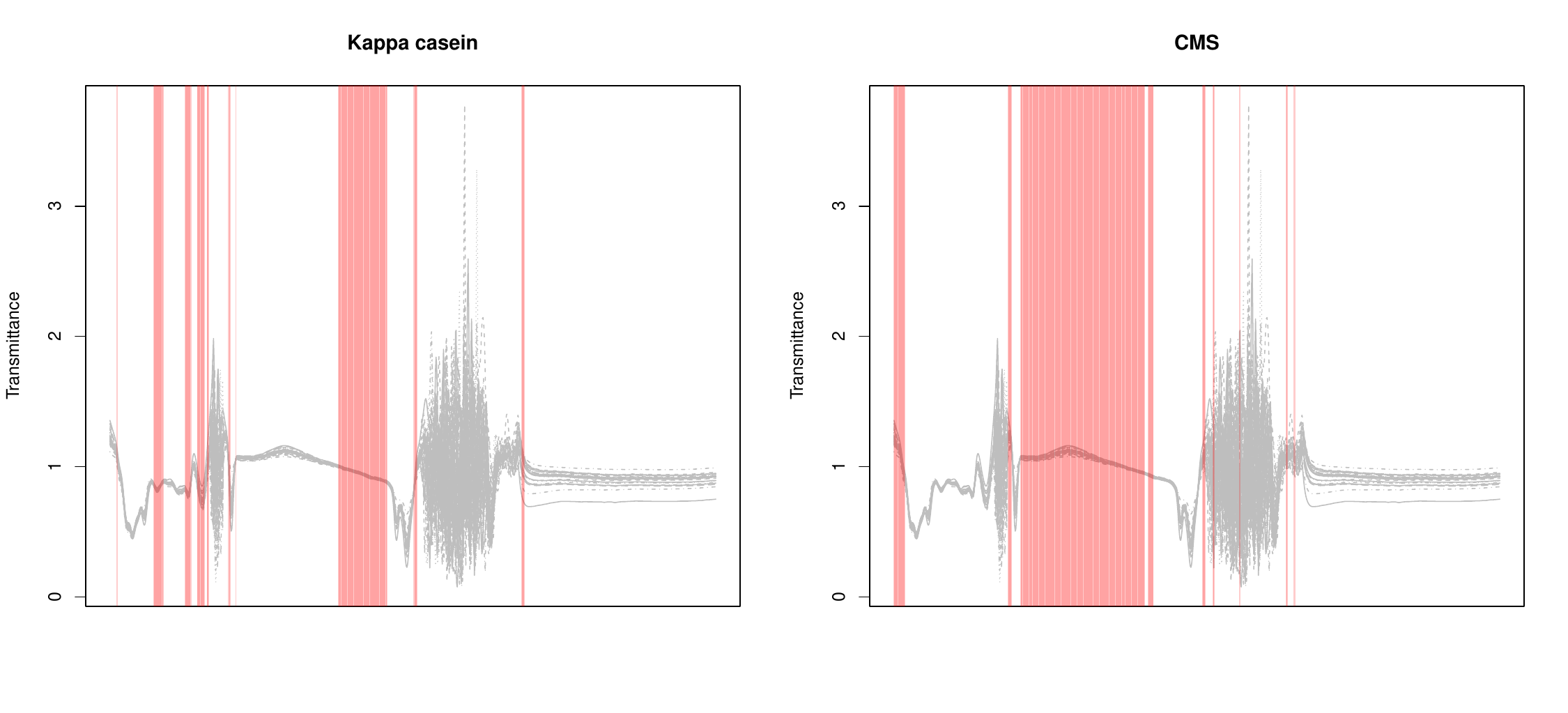} \\
    \vspace*{-1cm}
    \hspace*{-7.85cm} \includegraphics[width = 7.8cm, height = 7.5cm]{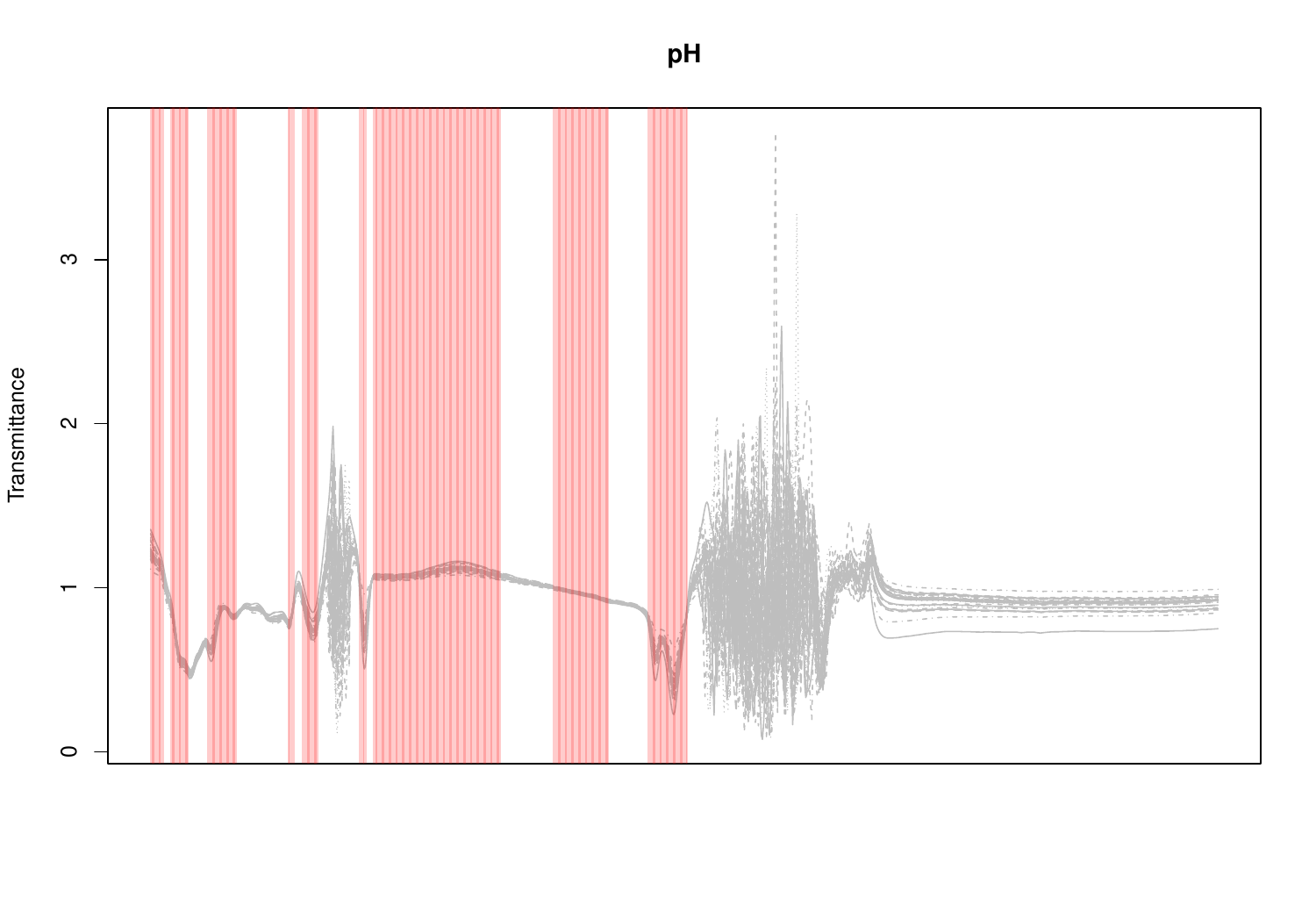} 
    \caption{Variables selected (red vertical lines) by S-PLS for the reference traits.}
    \label{fig:Agnieszka1}
\end{figure}

After removing the missing values for each trait, a calibration set was created based on the training data using the Puchwein algorithm implemented in the \texttt{prospectr} package \citep{prospectr} with the remaining samples left out for validation. The Puchwein algorithm selects samples by iteratively eliminating similar samples using the Mahalanobis distance. Even if no outliers have been removed beforehand, they have not been included in the calibration subset by the algorithm.

Afterwards principal component analysis was considered for mean-centered and MSC-treated mean-centered spectra. As an exploratory insight, it has been observed that a higher number of principal components was needed to capture 95\% of variance for MSC-treated spectra.

Several regression models were tested, including Random forest, Support Vector Machines, and sparse PLS (S-PLS). 
The models used to provide the test dataset predictions were the ones with the lowest RMSECV and with the best fit when considering a regression of the predicted values versus the reference ones. For $\kappa$-casein and pH, the best model obtained was S-PLS, implemented in the \texttt{spls} package \citep{splscit}, after removal of high-noise-levels (4 and 5 factors, respectively). For CMS the best results were obtained for the S-PLS model trained on the full spectrum after applying MSC (1 latent variable). The \texttt{caret} \citep{caret} training included three resampling iterations with 5-folds cross validation. 
Wavenumbers selected by the S-PLS model for each trait were plotted over the spectra in Figure \ref{fig:Agnieszka1} (311, 145, and 266 variables for pH, $\kappa$-casein, and CMS respectively). A noticeable overlap in variables selected has been observed between models predicting pH and CMS.


\subsection{Other approaches}\label{sec:sec3_8}
In the previous sections the best strategies, in terms of the obtained results, have been outlined and discussed. Nonetheless, some interesting approaches have been adopted also by other participants to the competition. In particular one of the participants detected the outliers by means of a multiple linear regression, fitted for each trait. Afterwards the outlying observations have been identified using the Bonferroni outlier test \citep[see e.g.][]{fox2015applied} and consequently deleted. Another researcher considered, when building the calibration models, the Interval Partial Least Squares model \citep[iPLS,][]{norgaard2000interval}. This is a graphically oriented method which can be seen as a local and interval-based extension of the standard PLS. As a consequence it might be particularly useful to interpret the results, since it highlights  different contributions from distinct spectral regions. Moreover a modification of the iPLS algorithm, called Backward Interval Partial Least Square \citep[biPLS,][]{norgaard2005multivariate}, has been tested. This approach works similarly to iPLS but it deletes sequentially the poorest performing interval. For an application of these algorithms on near-infrared data readers may refer to \citet{xiaobo2010genetic}.

\section{Discussion}\label{sec:sec4}

As mentioned in Section \ref{sec:sec2}, the contributions have been evaluated considering the RMSEP for each single trait while, for the overall performances, the relative error as defined in (\ref{eq:rel_err}) has been computed. By thoroughly looking at the final results, reported in Table \ref{table:final_results}, some relevant insights can be obtained. First of all, it is clear that the performances of different methods are usually trait specific. This seems coherent with the so called \emph{no free lunch theorem} \citep[see e.g.][]{wolpert1997no} which provides an essential warning about the danger of comparing and ranking different algorithms by studying their performances on a small set of prediction problems. A general conclusion to draw is concerned with the poor predictability of the CMS, thus providing coherent indications with the ones obtained in previous works \citep[see e.g.][]{visentin2015prediction,frizzarin_predicting_2021}. Note that different contributions performed comparably when predicting the test set values. Lastly, we would like to highlight that all the results need to be interpreted cautiously since the techniques have been evaluated on a small test set, making it difficult to properly generalize the findings. 

A common choice that all the participants had to make was concerned with the possibility to transform the data prior to running their analyses. A first possible transformation consisted in considering absorbance instead of transmittance values. The well known \emph{Beer-Lambert law} seems to suggest that working on the absorbance scale can lead to improvements in the performances, even if sometimes the benefits are minimal. The results in Section \ref{sec:sec3_1}, where the participants tried both the approaches without noting significant improvements, as well as prior experiences of some of the authors, seem to suggest that this transformation is not strictly needed when analyzing MIRS data. 

Other transformations have been applied on the spectra by some of the participants, as a pre-treatment step. Procedures such as the Multiplicative Scatter Correction or the Savitzky-Golay derivative smoothing have been tried but, again, without strong ameliorations of the results. Such transformations are usually hugely important when dealing with highly-noisy spectra, often encountered in the near-infrared spectroscopy applications. Nonetheless, in the dairy framework and when MIRS data are available, they seem to have a lower impact.

\begin{table}[t]
\centering
\caption{Root Mean Squared Error computed for the test set (RMSEP) for the reference traits and the overall relative error (RERR) as defined in \ref{eq:rel_err}. In bold, the best results are highlighted for each single trait and for the overall prediction. }
\begin{tabular}{l|ccc|c}
\hline
  Participant & & RMSEP & & RERR \\
   & $\kappa$-casein & CMS & pH &  \\ 
  \hline
  1 & 1.77 & 430.78 & {\bf 0.09} & 1.019 \\
  2 & 1.82 & 431.48 & 0.11 & 1.104 \\
  3 & 1.76 & 428.42 & 0.12 & 1.127 \\
  4 & {\bf 1.68} & 430.91 & {\bf 0.09} & {\bf 1.002} \\
  5 & 1.71 & {\bf 428.27} & 0.11 & 1.080 \\
  6 & 1.69 & 429.01 & 0.11 & 1.076 \\
   \hline
\end{tabular}
\label{table:final_results}
\end{table}

Another subjective decision that might have influenced the results was concerned with the outliers removal process. Outliers detection can be a critical step when working with spectroscopy data as it might be troublesome to discriminate between true extreme values and data collection errors, often due to the instruments used. In some of the contributions the analyses have been run considering all the available observations while some participants decided to follow the same approach considered in \citet{frizzarin_predicting_2021} where the observations have been considered as outliers when they had unusually large or small values for a reference trait; none of the participants considered procedures identifying outlying observations based on the spectra and therefore on unusual values of the wavenumbers. After the competition, the values for the traits on the test dataset were made publicly available. After re-running the analyses and re-evaluating their approach, a group of participants noted an improvement in the performances on the test set when outlying observations were not removed from the training set. This behavior might act as a partial confirmation of the challenges involved in the definition of the concept of outliers.

The participants were not provided with the specific information about the wavenumbers. Therefore, it was not straightforward to include prior information on the role of different wavenumbers in the modelling strategies. One of the possibly most impactful choices is concerned with the removal of the highly noisy regions related to water absorption processes, which are usually present when analyzing milk samples. In \citet{frizzarin_predicting_2021} these regions have been removed, according to the suggestions in \citet{hewavitharana1997fourier}. In the competition only two contributions considered the presence of the water absorption phenomenon: in fact in Section \ref{sec:sec3_4} this information has been used to devise an adaptive functional approach which provided good results both in terms of performances and in terms of interpretability. On the other hand, in Section \ref{sec:sec3_7}, the interquantile range for each wavenumbers has been used as a clever data-driven approach to identify noisy variables in the data. From previous experiences and from other analyses on similar data, it has been noted that the removal of highly noisy regions can be strongly influential and lead to improvements in the predictive performances. 

In the chemometrics framework, Partial Least Squares based methods still constitute the state of the art, as they are widely used in different applications usually providing good results. Nonetheless this challenge showed that it is worth looking at different modelling tools when analyzing spectroscopic data as they might slightly improve the quality of the obtained predictions or enhance the interpretability. In fact, one of the most interesting outcomes of the competition is related to the heterogeneity of the proposed approaches. Gathering together participants having different backgrounds, spanning from animal science to more statistical or machine learning ones, the contributions provided a wide overview about different possible ways to look at MIRS data. In most of the cases the spectra have been treated as standard vectors, also called \emph{tabular} approaches in Section \ref{sec:sec3_1}, thus disregarding the natural ordering of the wavenumbers. Nonetheless, the contributions in Section \ref{sec:sec3_2} and \ref{sec:sec3_4}, interestingly propose time series and functional approaches, respectively. The latter approach produced generally better predictions with respect to the first one; this might be due to the specific correlation structure usually encountered in MIRS data. In fact, even if neighboring wavenumbers tend to be highly correlated, strong relationships are often seen among spectral regions being far one from the other, since chemical constituents might have different absorption peaks at different spectral regions \citep[see][]{casa2021parsimonious}. Time series models, more often accounting for short term relationships and for more regular dependency patterns, might fail to properly capture this behaviour in spectroscopic data. On the other hand, the functional approach perfomed very well across the different traits, providing the lowest RERR in the challenge. Even if the spectra has already been treated before as continuous functions \citep{alsberg1993representation}, functional strategies are still not widespread in the chemometrics framework. Nonetheless, as pointed out in \citet{saeys2008potential} and as partially confirmed by the results of the competition, functional data analysis can find a fertile ground in infrared spectroscopy applications.  

A last relevant point, which have been discussed and faced both in the competition and during the whole workshop, is concerned with the adoption of methods that automatically perform wavelength selection. Often the participants advocated for the use of simpler methods which provide indications about the relevance of the features, therefore possibly providing more interpretable predictions from a practical standpoint. In fact such methods, jointly with proper collaborations with researchers with subject matter knowledge in the dairy framework, might lead to obtain information having a practical utility at different levels. For example from a commercial perspective, a proper identification of the most influential variables, can lead to the production of new spectrometers being cheaper and faster and possibly collecting data for those specific wavenumbers only. This can lead to a rapid and more pervasive diffusion of these technologies and, therefore, to an increase in the amount of data collected and to new challenges from a modelling standpoint. 

\section{Conclusion}\label{sec:conclusion}
The challenge organized during the first edition of the ``International Workshop on Spectroscopy and Chemometrics'' provided some insightful take home messages to both the participants and the researchers who attended to their presentations. First of all, the results showed how, when dealing with MIRS data, several different models provide good prediction and how their ranking can be strongly trait dependent. This advocates for the consideration of heterogeneous strategies when building calibration models in this framework. Moreover, it has been shown how different and less explored ways to treat spectral data, as for example adopting functional approaches, can lead to good results. Lastly, during the whole workshop, a lot of attention has been put on the necessity of relatively simple, parsimonious and interpretable models which can be more adequate when good predictions have to be complemented by the extraction of new knowledge about the phenomenon under study.

\section*{Acknowledgements}
This publication has emanated from research conducted with the financial support of Science Foundation Ireland (SFI) and the Department of Agriculture, Food and Marine on behalf of the Government of Ireland under grant number (16/RC/3835), the SFI Insight Research Centre under grant number (SFI/12/RC/2289\_P2) and the SFI Starting Investigator Research Grant ``Infrared spectroscopy analysis of milk as a low-cost solution to identify efficient and profitable dairy cows'' (18/SIRG/5562). Agnieszka Konkolewska is funded from the European Union's Horizon 2020 program (Marie Sklodowska-Curie grant agreement No. 841882).

\bibliographystyle{apalike}
\bibliography{biblio}

\begin{thebibliography}{}

\bibitem[Alsberg, 1993]{alsberg1993representation}
Alsberg, B.~K. (1993).
\newblock Representation of spectra by continuous functions.
\newblock {\em Journal of Chemometrics}, 7(3):177--193.

\bibitem[Bagnall et~al., 2017]{bagnall16bakeoff}
Bagnall, A., Lines, J., Bostrom, A., Large, J., and Keogh, E. (2017).
\newblock The great time series classification bake off: a review and
  experimental evaluation of recent algorithmic advances.
\newblock {\em Data Mining and Knowledge Discovery}, 31:606--660.

\bibitem[Bonfatti et~al., 2017]{bonfatti2017comparison}
Bonfatti, V., Tiezzi, F., Miglior, F., and Carnier, P. (2017).
\newblock Comparison of bayesian regression models and partial least squares
  regression for the development of infrared prediction equations.
\newblock {\em Journal of Dairy Science}, 100(9):7306--7319.

\bibitem[Casa et~al., 2021]{casa2021parsimonious}
Casa, A., O'Callaghan, T.~F., and Murphy, T.~B. (2021).
\newblock Parsimonious {B}ayesian {F}actor {A}nalysis for modelling latent
  structures in spectroscopy data.
\newblock {\em arXiv preprint arXiv:2101.12499}.

\bibitem[Cecchinato et~al., 2009]{cecchinato2009mid}
Cecchinato, A., De~Marchi, M., Gallo, L., Bittante, G., and Carnier, P. (2009).
\newblock Mid-infrared spectroscopy predictions as indicator traits in breeding
  programs for enhanced coagulation properties of milk.
\newblock {\em Journal of Dairy Science}, 92(10):5304--5313.

\bibitem[Chipman et~al., 2010]{bart}
Chipman, H.~A., George, E.~I., and McCulloch, R.~E. (2010).
\newblock {BART: Bayesian additive regression trees}.
\newblock {\em The Annals of Applied Statistics}, 4(1):266 -- 298.

\bibitem[Chollet, 2015]{chollet2015keras}
Chollet, F. (2015).
\newblock Keras.
\newblock \url{https://github.com/fchollet/keras}.

\bibitem[Chung et~al., 2019]{splscit}
Chung, D., Chun, H., and Keles, S. (2019).
\newblock {\em spls: Sparse Partial Least Squares (SPLS) Regression and
  Classification}.
\newblock R package version 2.2-3.

\bibitem[De~Boor, 1978]{de1978practical}
De~Boor, C. (1978).
\newblock {\em A practical guide to splines}.
\newblock Springer-Verlag, New York.

\bibitem[De~Marchi et~al., 2014]{de2014invited}
De~Marchi, M., Toffanin, V., Cassandro, M., and Penasa, M. (2014).
\newblock Invited review: Mid-infrared spectroscopy as phenotyping tool for
  milk traits.
\newblock {\em Journal of Dairy Science}, 97(3):1171--1186.

\bibitem[Dempster et~al., 2020]{dempster2019rocket}
Dempster, A., Petitjean, F., and Webb, G.~I. (2020).
\newblock {ROCKET:} exceptionally fast and accurate time series classification
  using random convolutional kernels.
\newblock {\em Data Mining and Knowledge Discovery}, 34(5):1454--1495.

\bibitem[Deng and Runger, 2012]{regrf}
Deng, H. and Runger, G. (2012).
\newblock Feature selection via regularized trees.
\newblock In {\em The 2012 International Joint Conference on Neural Networks
  (IJCNN)}, pages 1--8.

\bibitem[Durand, 2001]{durand2001local}
Durand, J.-F. (2001).
\newblock Local polynomial additive regression through pls and splines: Plss.
\newblock {\em Chemometrics and intelligent laboratory systems},
  58(2):235--246.

\bibitem[Fox, 2015]{fox2015applied}
Fox, J. (2015).
\newblock {\em Applied regression analysis and generalized linear models}.
\newblock Sage Publications.

\bibitem[Friedman et~al., 2010]{glmnet}
Friedman, J., Hastie, T., and Tibshirani, R. (2010).
\newblock Regularization paths for generalized linear models via coordinate
  descent.
\newblock {\em Journal of Statistical Software}, 33(1):1--22.

\bibitem[Friedman et~al., 2001]{friedman2001elements}
Friedman, J., Hastie, T., Tibshirani, R., et~al. (2001).
\newblock {\em The elements of statistical learning}.
\newblock Springer series in statistics, New York.

\bibitem[Frizzarin et~al., 2021]{frizzarin_predicting_2021}
Frizzarin, M., Gormley, I.~C., Berry, D.~P., Murphy, T.~B., Casa, A., Lynch,
  A., and McParland, S. (2021).
\newblock Predicting cow milk quality traits from routinely available milk
  spectra using statistical machine learning methods.
\newblock {\em Journal of Dairy Science}, 104(7):7438--7447.

\bibitem[Grelet et~al., 2015]{grelet2015standardization}
Grelet, C., Pierna, J.~F., Dardenne, P., Baeten, V., and Dehareng, F. (2015).
\newblock Standardization of milk mid-infrared spectra from a european dairy
  network.
\newblock {\em Journal of Dairy Science}, 98(4):2150--2160.

\bibitem[Hastie, 2017]{hastie2017generalized}
Hastie, T.~J. (2017).
\newblock {\em Generalized additive models}.
\newblock Routledge.

\bibitem[Hewavitharana and van Brakel, 1997]{hewavitharana1997fourier}
Hewavitharana, A.~K. and van Brakel, B. (1997).
\newblock Fourier transform infrared spectrometric method for the rapid
  determination of casein in raw milk.
\newblock {\em Analyst}, 122(7):701--704.

\bibitem[Ismail~Fawaz et~al., 2019]{IsmailFawaz2018deep}
Ismail~Fawaz, H., Forestier, G., Weber, J., Idoumghar, L., and Muller, P.-A.
  (2019).
\newblock Deep learning for time series classification: a review.
\newblock {\em Data Mining and Knowledge Discovery}, 33(4):917--963.

\bibitem[Kapelner and Bleich, 2016]{bartMachine}
Kapelner, A. and Bleich, J. (2016).
\newblock {bartMachine}: Machine learning with {B}ayesian additive regression
  trees.
\newblock {\em Journal of Statistical Software}, 70(4):1--40.

\bibitem[Karatzoglou et~al., 2004]{kernlab}
Karatzoglou, A., Smola, A., Hornik, K., and Zeileis, A. (2004).
\newblock kernlab -- an {S4} package for kernel methods in {R}.
\newblock {\em Journal of Statistical Software}, 11(9):1--20.

\bibitem[Kr{\"a}mer et~al., 2008]{kramer2008penalized}
Kr{\"a}mer, N., Boulesteix, A.-L., and Tutz, G. (2008).
\newblock Penalized partial least squares with applications to b-spline
  transformations and functional data.
\newblock {\em Chemometrics and Intelligent Laboratory Systems}, 94(1):60--69.

\bibitem[Kuhn, 2020]{caret}
Kuhn, M. (2020).
\newblock {\em caret: Classification and Regression Training}.
\newblock R package version 6.0-86.

\bibitem[McParland et~al., 2014]{mcparland2014mid}
McParland, S., Lewis, E., Kennedy, E., Moore, S., McCarthy, B., O’Donovan,
  M., Butler, S.~T., Pryce, J., and Berry, D.~P. (2014).
\newblock Mid-infrared spectrometry of milk as a predictor of energy intake and
  efficiency in lactating dairy cows.
\newblock {\em Journal of Dairy Science}, 97(9):5863--5871.

\bibitem[Mevik et~al., 2020]{pls}
Mevik, B.-H., Wehrens, R., and Liland, K.~H. (2020).
\newblock {\em pls: Partial Least Squares and Principal Component Regression}.
\newblock R package version 2.7-3.

\bibitem[Meyer et~al., 2021]{e1071}
Meyer, D., Dimitriadou, E., Hornik, K., Weingessel, A., and Leisch, F. (2021).
\newblock {\em e1071: Misc Functions of the Department of Statistics,
  Probability Theory Group (Formerly: E1071), TU Wien}.
\newblock R package version 1.7-6.

\bibitem[Nguyen et~al., 2019]{nguyen19interpretable}
Nguyen, T.~L., Gsponer, S., Ilie, I., O'Reilly, M., and Ifrim, G. (2019).
\newblock Interpretable time series classification using linear models and
  multi-resolution multi-domain symbolic representations.
\newblock {\em Data Mining and Knowledge Discovery}, 33(4):1183--1222.

\bibitem[Nguyen and Ifrim, 2021]{nguyen2021mrsqm}
Nguyen, T.~L. and Ifrim, G. (2021).
\newblock Mr{SQM}: Fast time series classification with symbolic
  representations.
\newblock {\em arXiv preprint arXiv:2109.01036}.

\bibitem[N{\o}rgaard et~al., 2005]{norgaard2005multivariate}
N{\o}rgaard, L., Hahn, M., Knudsen, L., Farhat, I., and Engelsen, S. (2005).
\newblock Multivariate near-infrared and raman spectroscopic quantifications of
  the crystallinity of lactose in whey permeate powder.
\newblock {\em International Dairy Journal}, 15(12):1261--1270.

\bibitem[N{\o}rgaard et~al., 2000]{norgaard2000interval}
N{\o}rgaard, L., Saudland, A., Wagner, J., Nielsen, J.~P., Munck, L., and
  Engelsen, S.~B. (2000).
\newblock Interval partial least-squares regression (i{PLS}): A comparative
  chemometric study with an example from near-infrared spectroscopy.
\newblock {\em Applied Spectroscopy}, 54(3):413--419.

\bibitem[Pedregosa et~al., 2011]{pedregosa2011scikit}
Pedregosa, F., Varoquaux, G., Gramfort, A., Michel, V., Thirion, B., Grisel,
  O., Blondel, M., Prettenhofer, P., Weiss, R., and Dubourg, V. (2011).
\newblock Scikit-learn: Machine learning in python.
\newblock {\em Journal of Machine Learning Research}, 12:2825--2830.

\bibitem[Pierna et~al., 2011]{pierna2011case}
Pierna, J. A.~F., Duval, H., Valderrama, P., Rutledge, D.~N., Baeten, V., and
  Dardenne, P. (2011).
\newblock A case study of extrapolation in {NIR} modelling—a chemometric
  challenge at ‘{C}himiom{\'e}trie 2009’.
\newblock {\em Chemometrics and Intelligent Laboratory Systems},
  106(2):205--209.

\bibitem[Pierna et~al., 2020]{pierna2020applicability}
Pierna, J.~F., Laborde, A., Lakhal, L., Lesnoff, M., Martin, M., Roggo, Y., and
  Dardenne, P. (2020).
\newblock The applicability of vibrational spectroscopy and multivariate
  analysis for the characterization of animal feed where the reference values
  do not follow a normal distribution: A new chemometric challenge posed at the
  ‘{C}himiom{\'e}trie 2019’congress.
\newblock {\em Chemometrics and Intelligent Laboratory Systems}, 202:104026.

\bibitem[{R Core Team}, 2020]{RCoreTeam2020}
{R Core Team} (2020).
\newblock {\em {R: A Language and Environment for Statistical Computing}}.
\newblock R Foundation for Statistical Computing, Vienna, Austria.

\bibitem[Ramsay and Silverman, 2005]{FDA}
Ramsay, J.~O. and Silverman, B.~W. (2005).
\newblock {\em Functional data analysis}.
\newblock Springer, New York.

\bibitem[Saeys et~al., 2008]{saeys2008potential}
Saeys, W., De~Ketelaere, B., and Darius, P. (2008).
\newblock Potential applications of functional data analysis in chemometrics.
\newblock {\em Journal of Chemometrics}, 22(5):335--344.

\bibitem[Stevens and Ramirez-Lopez, 2020]{prospectr}
Stevens, A. and Ramirez-Lopez, L. (2020).
\newblock {\em An introduction to the prospectr package}.
\newblock R package version 0.2.1.

\bibitem[Tan et~al., 2021]{Tan2020TSER}
Tan, C.~W., Bergmeir, C., Petitjean, F., and Webb, G.~I. (2021).
\newblock Time series extrinsic regression.
\newblock {\em Data Mining and Knowledge Discovery}, 35(3):1032--1060.

\bibitem[Van~Rossum and Drake, 2009]{python}
Van~Rossum, G. and Drake, F.~L. (2009).
\newblock {\em Python 3 Reference Manual}.
\newblock CreateSpace, Scotts Valley, CA.

\bibitem[Visentin et~al., 2015]{visentin2015prediction}
Visentin, G., McDermott, A., McParland, S., Berry, D.~P., Kenny, O., Brodkorb,
  A., Fenelon, M.~A., and De~Marchi, M. (2015).
\newblock Prediction of bovine milk technological traits from mid-infrared
  spectroscopy analysis in dairy cows.
\newblock {\em Journal of Dairy Science}, 98(9):6620--6629.

\bibitem[Wold et~al., 1987]{wold1987principal}
Wold, S., Esbensen, K., and Geladi, P. (1987).
\newblock Principal component analysis.
\newblock {\em Chemometrics and Intelligent Laboratory Systems}, 2(1-3):37--52.

\bibitem[Wolpert and Macready, 1997]{wolpert1997no}
Wolpert, D.~H. and Macready, W.~G. (1997).
\newblock No free lunch theorems for optimization.
\newblock {\em IEEE transactions on evolutionary computation}, 1(1):67--82.

\bibitem[Wright and Ziegler, 2017]{ranger}
Wright, M.~N. and Ziegler, A. (2017).
\newblock {ranger}: A fast implementation of random forests for high
  dimensional data in {C++} and {R}.
\newblock {\em Journal of Statistical Software}, 77(1):1--17.

\bibitem[Xiaobo et~al., 2010]{xiaobo2010genetic}
Xiaobo, Z., Jiewen, Z., Hanpin, M., Jiyong, S., Xiaopin, Y., and Yanxiao, L.
  (2010).
\newblock Genetic algorithm interval partial least squares regression combined
  successive projections algorithm for variable selection in near-infrared
  quantitative analysis of pigment in cucumber leaves.
\newblock {\em Applied spectroscopy}, 64(7):786--794.

\end{thebibliography}

\end{document}